# Pickering emulsions with α-cyclodextrin inclusions:

# Structure and thermal stability

Raul Diaz-Salmeron [a], Ismail Chaab [a,b], Florent Carn [b], Madeleine Djabourov [a*],

Kawthar Bouchemal [c]

a. Laboratoire de Physique Thermique, ESPCI-ParisTech, PSL Research University, 10 rue Vauquelin, 75231 Paris cedex 05, France.

b. Laboratoire Matière et Systèmes Complexes, UMR 7057 Université Denis Diderot-Paris 7, Case Courrier 7056, 10 Rue Alice Domon et Léonie Duquet, 75205 Paris, Cedex 13, France

c. Institut Universitaire de France, Institut Galien Paris Sud, UMR CNRS 8612, Université Paris-Sud, Faculté de Pharmacie, 5, rue J-B. Clément, 92296, Châtenay-Malabry cedex, France.

* Corresponding author: Madeleine.Djabourov@espci.fr

## Abstract

This paper explores structural, interfacial and thermal properties of two types of Pickering emulsions containing α-cyclodextrin inclusion complexes: on one hand, emulsions were obtained between aqueous solutions of α-cyclodextrin and different oils (fatty acids, olive oil, silicone oil) and on the other hand, emulsions were obtained between these oils, water and micro or nano-platelet suspensions with inclusion complexes of hydrophobically-modified polysaccharides. The emulsions exhibit versatile properties according to the molecular architecture of the oils. Experiments were performed by microcalorimetry, X-ray diffraction and confocal microscopy. The aptitude of oil molecules to be threaded in α-cyclodextrin cavity is a determining parameter in emulsification and thermal stability. The heat flow traces and images showed dissolution, cooperative melting and de-threading of inclusion complexes which take place progressively, ending at high temperatures, close or above 100°C. Another important feature observed in the emulsions with micro-platelets is the partial substitution of the guest molecules occurring at room temperature at the oil/water interfaces without dissolution, possibly by a diffusion mechanism of the oil. Accordingly, the dissolution and the cooperative melting temperatures of the inclusion crystals changed, showing marked differences upon the type of guest molecules. The enthalpies of dissolution of crystals were measured and compared with soluble inclusions.

# 1. Introduction

Stabilization of emulsion with colloidal particles is well known in food technology, cosmetics and pharmaceutical formulations. Nanoparticles and microparticles can be used as emulsifying agents which participate to the stabilization of the multi-phase structures by providing a physical barrier to the emulsion droplets coalescence. This type of stabilization is commonly called "Pickering stabilization" after the pioneering work of Pickering more than a century ago [1]. Depending on whether the particles are predominantly hydrophilic or hydrophobic, the systems may be either oil-in-water (O/W) or water-in-oil (W/O) emulsions. Ingredients commonly employed for making and stabilizing edible O/W emulsions are proteins and hydrocolloids [2–4]. The properties of Pickering emulsions [5] are usually determined by particle size [6], particle wettability [7], particle concentration [8], oil/water ratio [9], pH [10], salt concentration [11] and solvent type[12].

In addition to these classical stabilization modes, a particular mechanism of stabilization was identified for emulsions containing cyclodextrins in aqueous solutions mixed with oils. The complexation of cyclodextrins (CD) by "the threading mechanism" with poly(ethylene glycol) (pseudo polyrotaxane formation) was known since Harada's group publications [13–16] and besides, with a large number of other guest molecules. CD molecules in water have no surface activity. However, upon mixing $n$-alkanols and water containing various concentrations of $\alpha$-CD, Hashizaki et al [17] found that precipitated complexes formed at high CD concentrations and, when adsorbed to the oil water interface, served as an emulsifier, without addition of any surfactant. They analyzed the mechanism of emulsification and conclude that these emulsions were a "kind of Pickering emulsion". Indeed, the interfacial tension was markedly reduced with increasing $\alpha$-CD concentrations and it was found that this effect originates from the formation of a precipitated layer of the inclusion complex between oil and $\alpha$-CD, whereas the soluble inclusion complexes have little effect on surface activity. Only the precipitated complexes, adsorbed at the droplet surfaces, reduce the surface tension. The oil droplet sizes decreased with $\alpha$-CD concentration and it is shown that emulsion stability is closely correlated to the amount of precipitated complexes. Later, Inoue and co-workers investigated $n$-alkane water emulsions stabilized by $\alpha$, $\beta$ and $\gamma$ cyclodextrins [18]. Authors show that $\beta$ and $\gamma$ CD are more strongly adsorbed at the oil-water interface than $\alpha$-CD and they substantially decrease the interfacial tension. Another piece of evidence, with a different context, was provided by Sanemasa et al [19] who compared the complexation constants versus $n$ (2 to 12) of a series of surfactants, alcohols and alkanes and concluded that the complexation constants are much lower for alkanes than those of surfactants and alcohols and noticed that the hydrophilic polar groups ($SO_3^-$, $COO^-$ and $OH^-$) work as stabilizers rather than inhibitors and contribute through hydrogen bonding with CD molecules to enhance formation the complex.

In a series of papers Paunov and co-workers [20–22] examined the interfacial properties of CD based Pickering emulsions and developed experimental methods for visualizing the interfaces. Their work highlights in particular the effect of particle size of inclusion complexes and particle contact angles at interfaces in stability of Pickering emulsions.

In this work, we raise the question of the structure and thermal stability of autoassemblies composed of inclusion complexes in emulsions prepared with $\alpha$-CD dissolved at room temperature in aqueous solutions in three different cases: a) oil molecules can form inclusion complex with $\alpha$-CD; b) oil molecules cannot form inclusions with $\alpha$-CD; c) water phase contains micro-platelets, prepared according to Bouchemal and coworkers method [23,24]. These assemblies are generated by the interaction between $\alpha$-CD and hydrophobic alkyl chain grafted on polysaccharides. In these emulsions, the oil phase was type a) or b). The micro-platelets were obtained by mixing $O$-palmitoyl-dextran or $O$-palmitoyl-amylopectin with $\alpha$-CD in aqueous solutions.

Emulsions were prepared with various oils: hexanoic acid, octanoic acid, olive oil and silicone oil. The stability of the emulsions versus temperature was investigated by microcalorimetry. The structure of inclusion complexes was investigated by X-Ray diffraction techniques. The oil/water interface in emulsions containing platelets was visualized by confocal microscopy in transmission and reflection modes.

## 2. Materials

Hexanoic acid (>99.5%), octanoic acid (>98%), Sudan III and $\alpha$-CD (produced by Wacker Chemie AG, purity ≥ 98%) were from Sigma Aldrich (Saint Quentin-Fallavier, France). Olive oil was from Cooper (Melun, France) and silicone oil SV1000 from VWR (Fontenay-sous-Bois, France). Deionized water was from Milli®Q system with a resistivity of 18 MΩcm. The synthesis of O-palmitoyl-dextran and O-palmitoyl-amylopectin is reported in Supporting material.

## 3. Methods

### 3.1. Preparation of emulsions stabilized by $\alpha$-CD

$\alpha$-CD powder was dissolved in pure water (10wt%) by magnetic agitation at room temperature for 30 minutes, at 500 rpm until complete dissolution. To prepare emulsions, oil was added to $\alpha$-CD solution mainly with the ratio by weight 10/90 oil in water and intimate mixing was obtained by magnetic agitation during 24h or longer times if necessary, at 500 rpm allowing inclusion complexes to form or simply to observe dispersion of oil droplets in water phase.

With olive oil, three emulsion compositions were investigated: 10/90wt%, 25/75wt% and 40/60wt%. The emulsions were characterized by μDSC. After aging at room temperature for several days, the centrifugation of emulsions was performed with Eppendorf 5702RH centrifuge at 3000 rpm in falcon tubes during 30 min at 25°C in order to collect the sediment as a paste at the bottom of the tubes and to analyze the structure by X-Ray diffraction techniques.

### 3.2. Platelet preparation

Platelets were prepared by adding MilliQ® water in a glass vial containing O-palmitoyl-dextran or O-palmitoyl-amylopectin and $\alpha$-CD. The ratio of the concentration of O-palmitoyl-polysaccharide/ $\alpha$ -CD was 1/10wt%. Platelets were denoted O-palmitoyl-dextran/$\alpha$CD or O-palmitoyl-amylopectin respectively. Physicochemical characterization described in Supporting material shows that the self-assembled inclusions have mainly the shape of hexagonal, flat platelets. Their mean hydrodynamic diameter is in the range 2 to 2.6 μm.

### 3.3. Preparation of Pickering emulsions stabilized by micro-platelets

To prepare Pickering emulsions with micro-platelets, different oils were added to the aqueous suspension and magnetic agitation was allowed for 24h à 500 rpm to create the emulsion, previous to μDSC measurements. Mainly, emulsions with the composition 10/90 by wt were investigated.

### 3.4 Experimental techniques

MicroDSC (μDSC) experiments were performed with μDSC3evo from Setaram (Caluire, France) in batch Hastaloy cells by weighing emulsions with different compositions. The total weight of the samples was circa 0.8 g. The reference cell was filled with water. The temperature ramp started at 25°C after thermal equilibration and the heating was performed with a constant rate of +0.1°C/min to a final temperature varying between 100 and 120°C.

After centrifugation, the dense sediment was investigated by X ray diffraction. Patterns were recorded using a Panalytical Empyrean set-up equipped with a multi-channel detector (PIXcel 3D) using Cu-Kα radiation in the 3 – 50°range, with a 0.025° step size and 30s per step. The scattering vector ($q$) is defined as: $q = \frac{4\pi}{\lambda} \sin \theta$ where $\lambda$ is the wavelength and $\theta$ is the half of the scattering angle.

Confocal microscopy: 8-bits images were recorded with an inverted confocal microscope Leica TCS SP5 (from Leica Microsystems, Heidelberg, Germany) equipped with a Leica 63x (HCX Plan APO CS, NA 1.4) objective working distance 0.14 mm; the sampling rate is de 60 nm in x, y plane and 200 nm in z direction. The software used is Image J.

# 4. Results and Discussion

## 4.1    Emulsions of oils and α-CD aqueous solutions

α-CD solutions with a fixed concentration of 10wt%, were prepared first, and then the emulsification step was performed. A clear distinction could be made among the different oils used for emulsion preparation.

Silicone oil and α-CD aqueous solutions (oil/water 10/90wt%) let under magnetic agitation during 24h at room temperature show no emulsification at all: the two liquids remain separated by a clear interface, the denser (water) being at the bottom and the lighter (the oil) at the upper part of the vials. Okumura et al [25] noticed that inclusion complexes cannot be formed between α-CD and inorganic oil poly(dimethyl siloxane), (PDMS); the relative sizes of the cavities of α-CD to the cross-sectional area of the polymers being an important factor of the complex formation of polymers with cyclodextrins. Homogenous emulsification could be realized with hexanoic and octanoic acids, both oils are in the liquid state at room temperature, and with olive oil. These oils are poorly water soluble. Hexanoic acid and octanoic acid emulsions were O/W with 10/90 by weight and with olive oil the volume fraction O/W was varied: 10/90, 25/75 and 40/60wt%. The hexanoic acid emulsions were let under magnetic agitation during variable periods of time. Also, the emulsions were centrifuged and the bottom was collected and analyzed by X-Ray diffraction.

### 4.1.1  Hexanoic Acid

The growth of inclusion complexes formed between α-CD and hexanoic acid is evidenced first by visual observation. When oil is added to the aqueous solution a white precipitate starts to form immediately at the interface and grows in the aqueous phase. The precipitate is an inclusion of oil into α-CD. Indeed, the width of the hydrophobic part of the guest has to be lower than 4.5 Å to allow formation of a stable inclusion compound with α-CD. This is the case with linear alkyl chains such as hexanoic acid, with a cross-section which does not exceed 4.5 Å in the zig-zag conformation. The emulsion was let 30 min under agitation. The oil contains a red color dye (Sudan III) to help the visualization. When agitation stops the vial shows a white phase and a (pink) layer of oil on the top of it, as is shown in figure 1. The oil was not included completely in the emulsion phase.

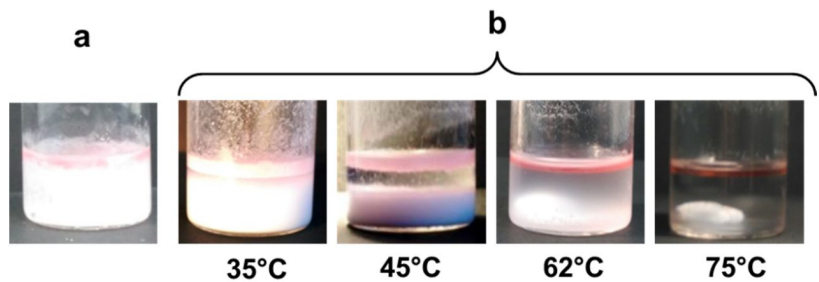

a                        b

35°C          45°C          62°C          75°C

**Figure 1** (a) Emulsion between hexanoic acid and water solution containing 10wt.% α-CD after 30 min agitation at room temperature. The emulsion is a white phase and the oil is a pink fluid layer on the top. (b) The vial was put in the oven at different temperatures, up to 75°C.

The vial was put into an oven, at increasing temperatures between 35 and 75°C, without any agitation and it was clearly seen that the emulsion progressively dissolves and finally the two phases, water and oil, totally separate as clear phases at 75°C.

Emulsions were more precisely investigated by μDSC after 30 minutes, 3 days and 5 days emulsification under stirring by a magnetic agitator. The heat flow traces from μDSC are shown Figure 2.

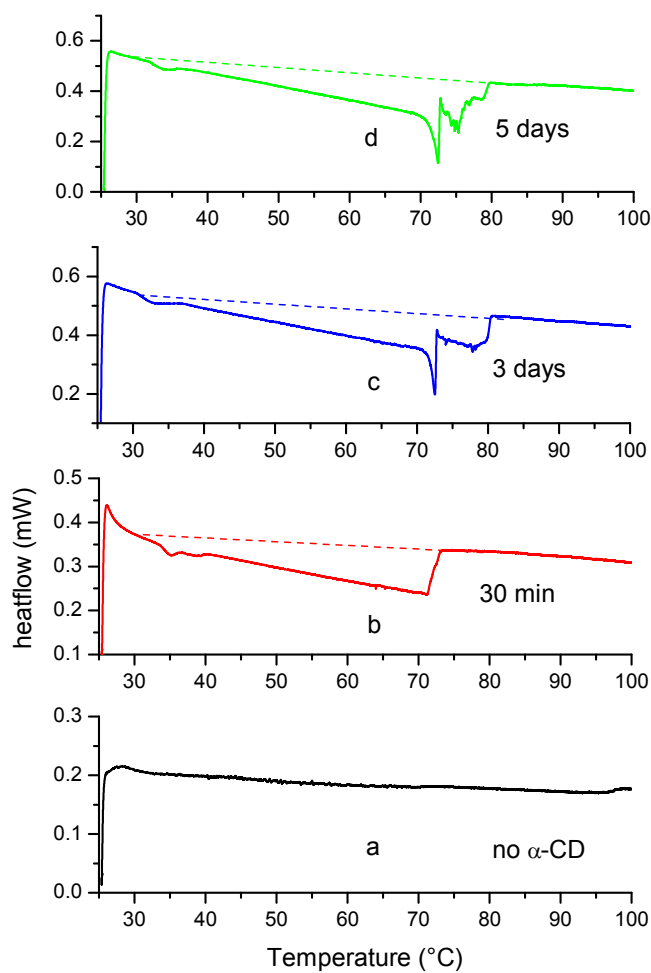

**Figure 2** Thermal behavior of inclusion complexes formed in emulsions between hexanoic acid and aqueous solutions: a) no α-CD in the water; b) The aqueous solution contains α-CD, (10 wt.%) and was mixed with hexanoic acid 30 min before measuring; c) After 3 days mixing, more stable inclusions are formed and a distinct crystalline phase appears which melts at the temperature $T_m$=72.7°C; d) After 5 days emulsification, the melting peak position is unchanged. The total area under the base line increases. The dotted lines indicate the base lines of the heat flow.

From bottom to the top: a) water and hexanoic acid are mixed, without α-CD, there is no heat exchange between the phases, they are completely immiscible; b) The aqueous solution contains α-CD and when mixed with hexanoic acid, an emulsion forms as shown in figure 1. The heat flow trace shows that the inclusions dissolve with increasing temperatures. The oil is progressively released at the top of the emulsion after de-threading of the α-CD inclusions, coalescence of oil droplets and creaming by density effects. The emulsion prepared after 30 min mixing is completely dissolved below 73°C.The heat flow returns to the base line at 73°C. The area under the base line, between 30 and 80°C is 1.6 J in this example. c) After 3 days mixing, more stable inclusions are formed and a distinct crystalline phase appears, built of α-CD inclusions, which melts at the temperature $T_m$= 72.7°C, near to the limit temperature of dissolution in previous case. The melting peak is very narrow (±1°C) and it is followed by an irregular endothermic signal which ends around 80°C. The cooperative melting releases α-CD molecules and destabilizes the hexanoic acid inclusions. Hexanoic acid molecules assemble or coalesce and "cream" towards the top of the cell, creating relative displacements of the two fluids, the oil and the aqueous solution, between 72°C and 80°C. An additional heat contribution appears after the melting peak in relation with the droplets formation. The total area under the base line is now 2.4 J for the same composition as case b). d) The same conclusions are confirmed after 5 days emulsification. The melting peak position is unchanged. The total area under the base line is 2.6 J. A large part of the oil inside inclusion complexes is released by heating, starting around 30°C. Progressive de-threading is a consequence of dissolution of the small crystallites, before the cooperative melting transition of the largest crystals, which is observed at 72°C. The curves suggest that the dissolution of inclusions increases with temperature. Melting transition of crystalline inclusions is a distinct process at a fixed temperature and is associated with the long time maturation of the crystallites. All hexanoic acid inclusions were released during the heating ramp, below 80°C.

### 4.1.2 Octanoic acid

A visual observation of vials containing this emulsion is shown Figure 3. The red dye (Sudan III) was added into the oil. Figure 3 shows images of the vial prepared *without any agitation*: a rapid growth is observed of the inclusion complex at the interface, at room temperature. Dense and large crystals are formed. They grow and sediment at the bottom of the vial. After 10 days at 25°C the vial shows three distinct phases: at the bottom, the crystals have settled as a translucent, slightly colored deposit, on the top of it, the intermediate phase is a transparent liquid -the aqueous phase-, and liquid layer at the surface (dark pink) is the oil in excess. The vial was put into the oven at different increasing temperatures. The inclusion complex crystals are very stable and very few dissolve until 75°C. At 100°C the large crystals at the bottom start melting, but without shaking to homogenize the solution in the vial, the process seems to be very slow. However, after shaking rapidly the vial, the last crystals lying at the bottom, finally dissolved. There were no more crystals left at 100°C.

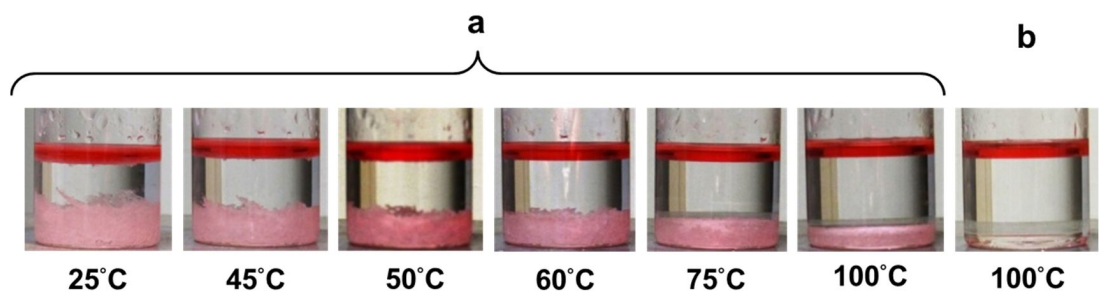

**Figure 3** Octanoic acid inclusions formed at the interface between water and octanoic acid without agitation (a); after shaking rapidly by hand at 100°C (b). The vial was put in the oven at different temperatures, up to 100°C. The oil phase is the pink color; the inclusion complex form large, macroscopic crystals which sediment at the bottom.

Images of octanoic acid emulsions prepared by agitation are shown in Supporting material section. Emulsions were investigated by μDSC after magnetic agitation during 3 days. The heat flow trace of the emulsion (10/90wt%O/W) in Figure 4 shows a sharp endothermic peak due to melting of the inclusions assembled in crystalline structures at $T_m$=92°C. The melting temperature of α-CD in water in supersaturated solutions (concentrations >50wt %) is $T_m$=76.5 ±0.5°C [26]. It is remarkable that the melting temperature of octanoic acid inclusions is above both melting temperatures of individual components, the α-CD crystalline hydrates (76.5°C) and the octanoic acid pure crystals (17°C). The dash line in Figure 4 indicates the base line of the heat flow. Because of the high melting temperature, the end of the dissolution is not fully accomplished and the final base line is not visible. The crystals of octanoic acid inclusions have a marked thermal stability and melt cooperatively. The melting entails necessarily the disruption of hydrogen bonds of the α-CD crystals and de-threading of the oil molecules. The surface under the base line, shown in Figure 4, between 30°C and 93°C is 2.86 J. The total enthalpy under the base line corresponds to progressive dissolution and to melting of the inclusion complex crystals, possibly of various sizes. The full dissolution is not achieved at 92°C and estimation of the additional contribution of the non-dissolved crystals from the µDSC data cannot be done.

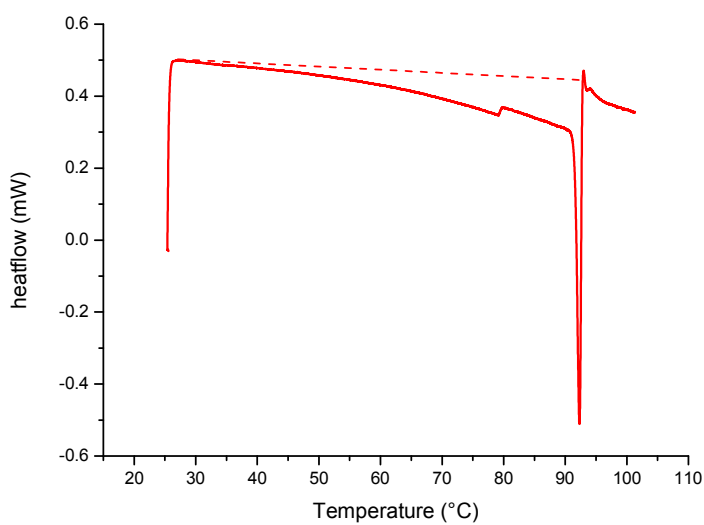

**Figure 4** Thermal behavior of inclusion complexes formed in emulsions between octanoic acid and α-CD aqueous solutions (10wt%) (3 days under agitation before measuring). The heat flow trace shows a clear melting peak of the crystalline inclusion complexes between α-CD and octanoic acid at $T_m$=92°C.

### 4.1.3 Olive oil

Vegetable oils are composed of triglycerides together with di- and monoglycerides, as well as free fatty acids. Olive oil contains a mixture of fatty acids varying from about 14 to 24 carbon atoms in length. The main fatty acids in olive oil are triacylglycerols (oleic acid, linoleic, palmitic, stearic acid and linoleic acid in various proportions), most prevalent being the oleic-oleic-oleic triacylglycerol. The vegetable oils can make inclusions with α-CD molecules [27,28].

The inclusion complexes with olive oil were prepared at different O/W ratios: 10/90, 25/75 and 40/60wt% with emulsification during 24h at 500 rpm. The heat flow traces are shown in Figure 5. Endothermic heat flow signals appear with a broad thermal evolution. The base line is chosen for simplicity as a straight line between 65°C and 95°C. From top to bottom: a) The O/W 10/90wt% shows a dissolution profile which starts around 65°C. The endothermal heat flow is 0.3 J. The measurement shows a broad thermal range of dissolution; b) the 25/75wt% O/W emulsion exhibits an even broader distribution: the dissolution starts around 30°C; below 60°C is the first population of complexes, followed by a second dissolution wave between 60 and 95°C. The total area is 1.38 J approximately with a base line as indicated on the figure. One can conclude that a larger amount of oil enables more inclusions to form spontaneously during the mixing stage at room temperature, but some of the inclusions are not as stable as in the case 10/90wt%; c) with more olive oil, 40/60wt%., the dissolution peak is larger 1.5 J and broadens continuously between 30°C and 90°C. The heat flow profiles indicate a progressive dissolution of the inclusion complexes and no melting temperature of well-structured crystals, as it was seen in hexanoic acid and octanoic acid complexes. The composition of olive oil containing various proportions of fatty acids and triglycerides does not allow the crystals to grow in size.

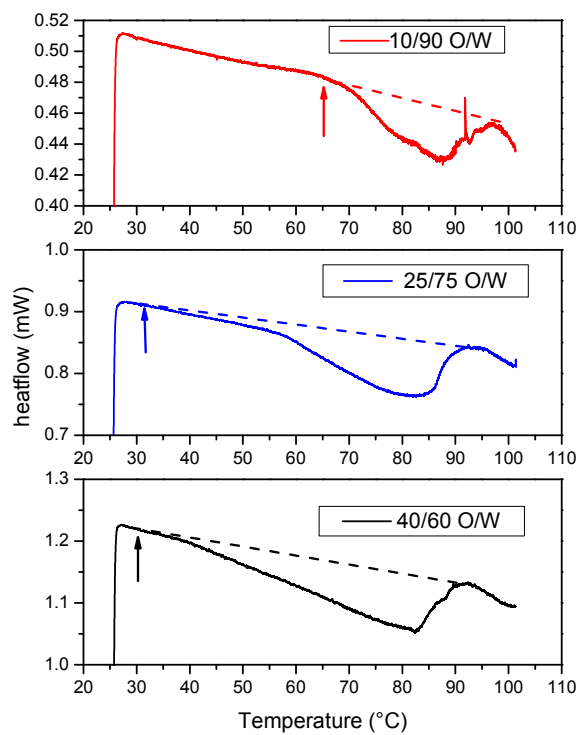

**Figure 5** Heat flow traces of emulsions formed between olive oil and water containing 10wt% α-CD. The O/W fraction was varied by weight from top to bottom: 10/90; 25/75; 40/60wt%. Emulsions were magnetically agitated during 24h prior to μDSC experiments.

## 4.2    X ray diffraction of centrifuged emulsions

The emulsions, after centrifugation at room temperature, are separated into a dense paste containing the inclusion complexes and a supernatant as a turbid liquid phase. X-ray diffraction performed on the paste collected at the bottom of the centrifugation tubes show for the three emulsions, distinct diffraction peaks associated with a crystalline structure (Figure 6). The positions of the diffraction peaks are similar in the three pastes which indicate an identical spacing between α-CD molecules in small crystallites. This suggests that the absence of a definite melting temperature in μDSC experiments, like in olive oil, does not mean that inclusion complexes do not make crystals in the emulsion, but only small crystals which do not reach the critical size, where the melting temperature is identified. The small crystals dissolve during the heating ramp.

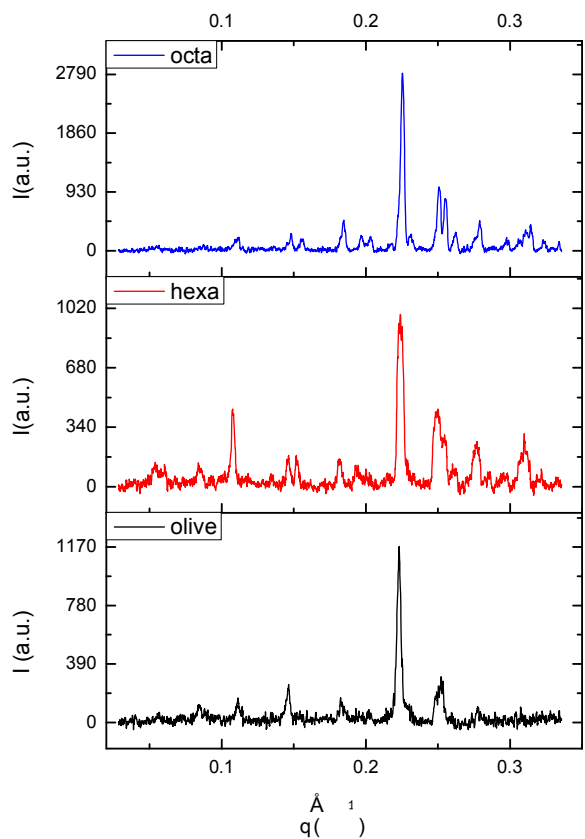

**Figure 6** X ray diffraction patterns of inclusion complexes collected after centrifugation of the emulsions with different oils. From top to bottom: octanoic acid, hexanoic acid and olive oil inclusion complexes with α-CD.

The X ray diffraction diagrams are different from the diffraction pattern of α-CD crystals in the "cage" structure. It is expected that they correspond to a "columnar structure" whose crystallographic parameters are not determined yet.

### 4.3   Discussion and interpretation

Previous works on heat of mixing with inclusion formation have been performed on binary solutions containing the solutes at *very high dilutions* [29–31]. The authors determined the thermodynamic parameters for the association between α-CD and monocarboxylic acids at pH=1.3 and pH=11.3, in order to show the role of the hydroxyl group in the complex formation and to compare the charged and the uncharged carboxyl groups. Monocarboxylic acids were investigated from $C_3$ to $C_9$ at pH=1.3 and up to $C_{12}$ at pH=11.3. The enthalpies of inclusion formation are negative and vary with the alkyl chain length. Enthalpies are the sum of several contrasting effects but dominated by the decrease in energy when a hydrophobic residue fills the cavity. The stoichiometry of the complexes with monocarboxylic acids was assumed to be 1:1. It was found: at pH=1.3, ΔH=-26.5kJ/mole for hexanoic acid and -42 kJ/mole for octanoic acid and at pH=11.3, ΔH=-6.1kJ/mole for hexanoic acid and -18.6 kJ/mole for octanoic acid. Monocarboxylic acids at pH=1.3 interact with α-CD with constants and enthalpies greater than those at pH=11.3. The interaction is enthalpy-driven for all acids.

Inclusion complexes with hexanoic acid and octanoic acid prepared in this work with 10/90 wt% O/W emulsions contain by wt. 10% α-CD molecules in the water phase. Considering that the samples in the μDSC cells contain 0.8g in total, there is 0.072g of $\alpha$-CD in each sample. As $M_w$=973 g/mole, the emulsions in the measuring cells contain $7.3\ 10^{-5}$ moles of $\alpha$-CD.

Emulsions with the composition 10/90wt%, contain $6.88\ 10^{-4}$ moles of hexanoic acid and $5.54\ 10^{-4}$ moles of octanoic acid. Therefore, with the assumption that complexes have a stoichiometry 1:1 the limiting factor for inclusion formation is the number of $\alpha$-CD molecules, while the oil molecules are in excess by a factor of 10. Upon heating, the maximum enthalpy of dissolution or melting and de-threading of inclusions is controlled by the number of $\alpha$-CD molecules forming the complexes. It is shown in the present work that the inclusion complexes are small crystals that grow in size and precipitate. The composition of the aqueous phase contains less and less $\alpha$-CD molecules and the number of inclusion complexes increases with time, sometimes without agitation. One can suppose that, if the process came to completion, the majority of $\alpha$-CD molecules formed inclusions with a large fraction in the crystalline structure. The total enthalpy measured by μDSC is due to dissolution and melting of the crystals and de-threading. Table 1 summarizes the measurements for hexanoic acid, octanoic acid and olive oil.

| Hexanoic acid emulsion | | Octanoic acid emulsion | Olive oil emulsion | |
|---|---|---|---|---|
| 30 min | 1.55 ± 0.10 J | - | 10/90 | 0.31 ± 0.10 J |
| 3 days | 2.38 ± 0.10 J | 2.86 ± 0.10 J | 25/75 | 1.38 ± 0.10 J |
| 5 days | 2.63 ± 0.10 J | - | 40/60 | 1.57 ±0.10 J |
| $|\Delta H|$ /mole $\alpha$-CD | **39 ± 1 kJ/mole** | **>36 ± 1 kJ/mole** | **Max ~32 ± 1 kJ/mole** | |

**Table 1.** Enthalpies of dissolution or melting and de-threading of inclusions self-assembled in crystalline forms with different oils.

The enthalpies of dissolution/de-threading per mole of a complex between $\alpha$-CD and hexanoic acid or octanoic acid are much larger than the values determined by Castronovo et al [29–31]. These values can be also compared to the enthalpies of dissolution of $\alpha$-CD hydrates in the "cage structure": 37.4 kJ/mole and for the cooperative melting of the crystals of $\alpha$-CD in supersaturated solutions, 20 kJ/mole [26].

The maximum enthalpy determined in this work for hexanoic acid inclusions, $|\Delta H|$=39 ±1 kJ/mole, is the sum of several contributions: the dissolution enthalpy of complexes between 30°C and 70°C, then the melting enthalpy of larger crystals with the narrow peak around 73°C and last, between 73°C and 80°C, the additional heat flow due to the complete dissolution of inclusions and the transfer of oil droplets to the surface. These processes include: a) the breaking of the hydrogen bonds stabilizing crystals of various sizes, which dissolve progressively at different temperatures, b) the enthalpy of de-threading of the oil molecules out of the CD cavity and c) further on, aggregation of the hydrophobic portions and formation of droplets which rise to the surface, towards the oil phase.

The dissolution enthalpy for octanoic acid complexes is larger after 3 days emulsification and shows a large cooperative melting. The crystals are more stable with temperature and the enthalpy of dissolution/melting is probably larger because the process could not be measured at the very high temperatures and because the alkyl groups are longer and enthalpies of association are probably larger. For a more precise evaluation of the enthalpy of dissolution of the self-assemblies of inclusion complexes one should vary the $\alpha$-CD concentration and the oil fraction. The high melting temperature of octanoic acid inclusions is a limiting factor for these experiments for determining the total enthalpy.

In olive oil emulsions the dissolution enthalpies are slightly smaller and show that the available $\alpha$-CD molecules did not all fully participate to inclusion complex formation because of the poly molecular composition of the oil.

# 5. Pickering emulsions with micro-platelets in aqueous dispersions

This section explores the behavior of small inclusion complexes self-assembled in supramolecular platelets, dispersed in aqueous medium and investigates the time and thermal stability of this type of Pickering emulsions. The micro and nano-platelets appear on TEM images as flat particles with a hexagonal shape for the largest ones, with however a broad size distribution of their size from 500 nm up to several microns. The suspension where the platelets were prepared is a white opaque liquid, with sometimes a deposit of the largest particles at the bottom; the density of the inclusion crystals made of cyclodextrins is large and if the crystals grow in size beyond 1μm they naturally sediment under gravity. This limitation was also observed in the hexanoic acid and octanoic acid inclusions reported in Section 4.

Emulsions were prepared by mixing the initial suspension of platelets with oil (hexanoic acid, oleic acid and silicone oil) during 24h under agitation. Emulsions were prepared with various O/W ratios and also after dilution of the aqueous suspensions by a factor of 10.

## 5.1 Pickering emulsions with micro-platelets and hexanoic acid

Comparison is established between emulsions prepared with hexanoic acid and various types of micro platelets on one hand, and emulsions formed directly between aqueous solution of α-CD and hexanoic acid. Typically emulsions have proportions fixed to 10/90wt% for O/W. The experiments are performed by μDSC.

The results are shown in Figure 7 for three cases: a) platelets with inclusions type *O*-palmitoyl-dextran/α-CD in water; b) platelets with inclusions *O*-palmitoyl-dextran/α-CD in water and in emulsion with hexanoic acid at room temperature and c) emulsion between α-CD (10% by wt.) dissolved in water and hexanoic acid, after mixing during a short time (30 min) to form a fresh emulsion. The heat flow traces, shown in Figure 7 are divided in two categories: the platelets in water have a very high stability and exhibit a cooperative melting at $T_m$=104°C with a sharp melting peak. The platelets *O*-palmitoyl-dextran/α-CD are in their initial preparation suspension. The platelets Figure 7a show an exceptional thermal stability. However, when they were emulsified with hexanoic acid at room temperature under mild agitation, the inclusions were fully substituted, as it is seen in Figure 7b: during the heating ramp, inclusion complexes exhibit a flow trace similar to the emulsions shown in Figure 7c prepared with hexanoic acid and α-CD in water. These experiments highlight the large difference between the thermal stability of the platelets in water and their natural tendency to exchange guest molecules with other ones, like hexanoic acid, the exchange occurring at room temperature by simply mixing during 30min with this particular oil. The substitution of inclusions occurs despite of crystalline structure of the platelets, at room temperature, without going through dissolution of the platelets, possibly through a molecular diffusion mechanism. The melting temperature of the "as prepared platelets" in water is associated to the cooperative disruption of their crystalline structure and the subsequent release of their inclusions by de-threading. The areas under the base lines are 1.17 J for *O*-palmitoyl-dextran/α-CD water between 30°C and 104°C, and 1.6 J for emulsion between hexanoic acid water and α-CD. The heat flow area of *O*-palmitoyl-dextran/α-CD in water is smaller and shows that the dissolution is not fully accomplished at 104°C because the heat flow does not return to the base line after the melting peak. The dissolution of the substituted platelets between 27°C and 85 °C gives an enthalpy of 1.1 J with a broad thermal distribution.

Considering platelets which can be used with therapeutic applications, at the body temperature, the fact that substitution of inclusions is likely to take place spontaneously is an important observation. Other guest molecules may exchange with those present at a certain time in the platelets. Exchanges between guest molecules in α-CD inclusions could be influenced by other complexes present in the formulation or by additional molecules belonging to the biological environment.

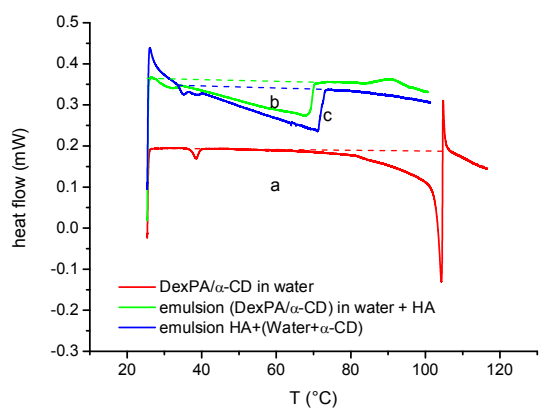

**Figure 7** Heat flows: a) (red line) Platelets composed of *O*-palmitoyl-dextran/α-CD inclusion complexes (dexPA/α-CD) in water; b) (green line) platelets in water and emulsion with hexanoic acid (HA); c) (blue line) emulsion between hexanoic acid and α-CD in water. The platelets in emulsion with hexanoic acid lost their thermal stability.

## 5.2 Pickering emulsions with micro-platelets and octanoic acid

Emulsions with micro platelets in water and octanoic acid were prepared at room temperature. The emulsions were introduced in the µDSC cells and submitted to the heating ramp, like in previous experiments. Figure 8 shows the heat flows for *O*-palmitoyl-dextran/α-CD platelets in emulsion with octanoic acid. The heat flow profile for *O*-palmitoyl-dextran/α-CD in emulsion with octanoic acid changed by comparison with Figure 7c when the oil phase was hexanoic acid. Inclusions were substituted with octanoic acid and the complexes dissolve progressively until 120°C with a major step change of the heat flow around 96°C and a subsequent peak around 102°C. We assume that these inclusions are no more *O*-palmitoyl-dextran/α-CD, but the alkyl inclusions were partially or totally replaced by octanoic acid molecules which provide a much more stable sequence because of the length of the alkyl chain.

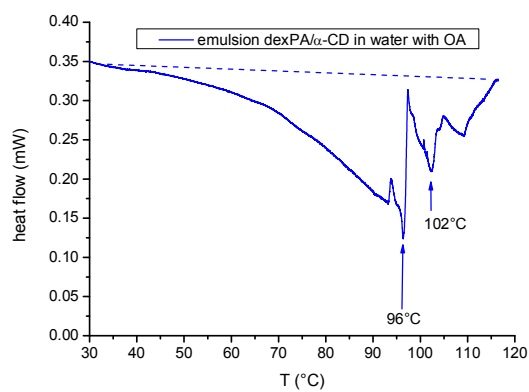

**Figure 8** Heat flow trace for *O*-palmitoyl-dextran/α-CD micro-platelets (dexPA/α-CD) in water, emulsified with octanoic acid (OA). The heat flow is totally changed in comparison with Figure 7.

For comparison, octanoic acid and α-CD inclusions form crystals melting at $T_m$= 92°C (figure 4). By molecular substitution of inclusions the stability of the complex is weakened. The dissolution starts around 30°C and secondary peaks are observed until 120°C. The substitution of the guest molecule was not fully accomplished in this example. The total area under the peak between 30 and 115.5 °C is 2.86 J. If all the α-CD inclusions in platelets have been substituted with octanoic acid, the total dissolution enthalpy of substituted inclusions would be 39kJ/mole of α-CD which is the same order of magnitude as reported in Table 1. The substitution of guest molecules in the platelets preserves the supramolecular assemblies of α-CD molecules but modifies the thermal stability of the structures.

## 5.3    Pickering emulsions with micro-platelets and silicone oil

In this section, we consider the interaction between silicone oil and platelet suspensions in water. We noticed that Pickering emulsions can be easily prepared. The difference between the emulsions with carboxylic acids and those with silicone oil, while the water phase contains α-CD molecules, is first of all that the silicone molecules are not able to thread inside the α-CD cavities because their diameter is too large to penetrate inside the cavity. Pickering emulsions were investigated by confocal microscopy and µDSC.

### 5.3.1    Confocal microscopy

We were interested in the emulsifying properties of the platelets, with silicone oil. The stabilization of Pickering emulsions by nano-platelets is less well known compared to spherical or needle like particles at droplet interfaces [21,22].

Images of Pickering emulsions were taken by confocal microscopy with platelets of *O*-palmitoyl-dextran/α-CD and *O*-palmitoyl-amylopectin/α-CD.

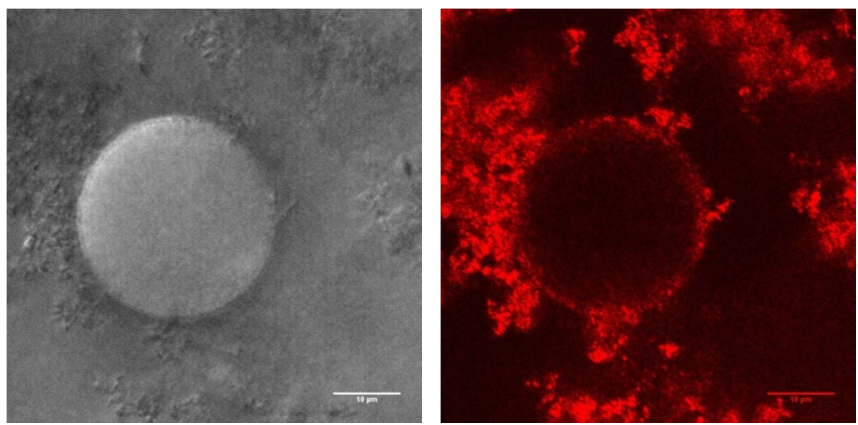

**Figure 9** Platelets of *O*-palmitoyl-dextran/α-CD in undiluted suspension, in emulsion 25/75wt% with silicone oil. Left: transmission mode; right reflection mode (bar=10µm).

Figure 9 is taken with *O*-palmitoyl-dextran/α-CD in undiluted suspension, 25/75wt% O/W. The surface of the droplet attracts the platelets and is covered with the small platelets. Although the platelets are dispersed in water, they tend to aggregate when the suspensions are kept without agitation. In this image they form a branched network which entraps the oil droplets. The observation by the reflection mode illustrates the crystalline nature of the platelets.

It was decided to dilute 10 times the initial suspension. The next images are taken with *O*-palmitoyl-amylopectin/α-CD after the 10 times dilution of the aqueous suspensions and emulsification with silicone oil observed in transmission and reflection modes of confocal microscopy.

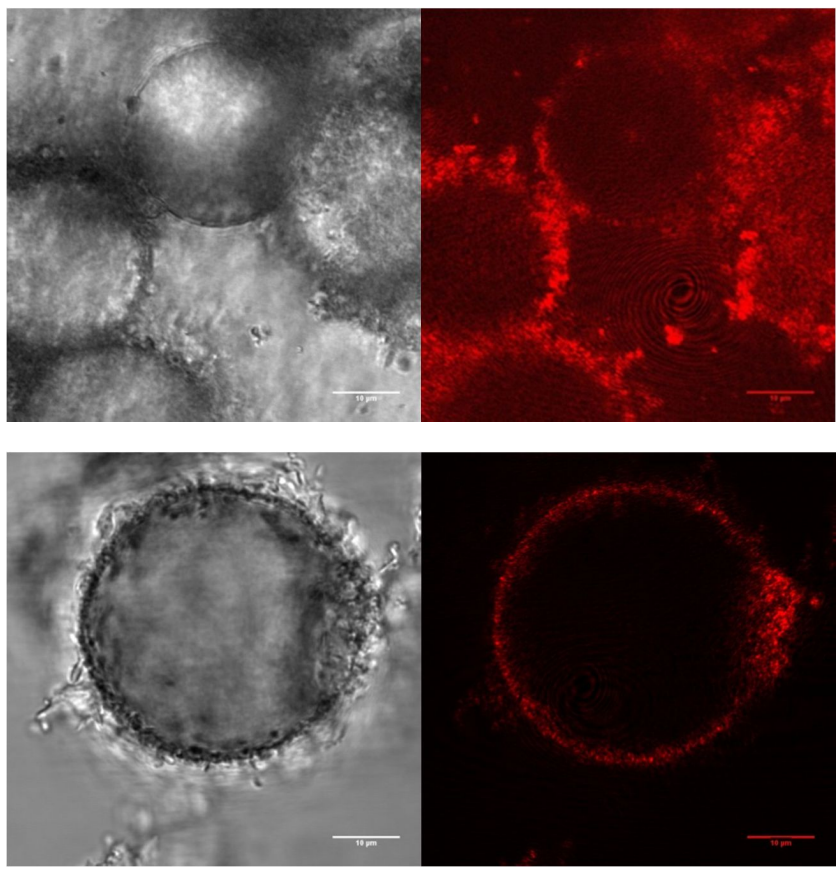

**Figure 10** Platelets of *O*-palmitoyl-amylopectin/α-CD diluted 10 times in emulsions with 10/90wt.% silicone oil in transmission (left) and reflection (right) modes (bar = 10μm).

The pictures, Figure 10 show several droplets in close contact covered with small platelets. The image in refection shows the thickness of the layer adsorbed at the surface of the droplets between 1 and 5 μm. By dilution, the platelets are not present anymore in the surrounding water phase, only at the surface of the oil droplets. The droplets in close contact do not coalesce because the layer of platelets creates an efficient barrier. The bottom images show a thin layer of small platelets around one droplet, without any free platelet in the surrounding liquid (water phase). Other pictures of the silicone oil droplets can be seen in Supporting material section.

### 5.3.2 Thermal stability of micro-platelets by μDSC

The thermal stability of the platelets was investigated by μDSC in emulsion and compared with the thermal stability in water. Results are shown in Figure 11.

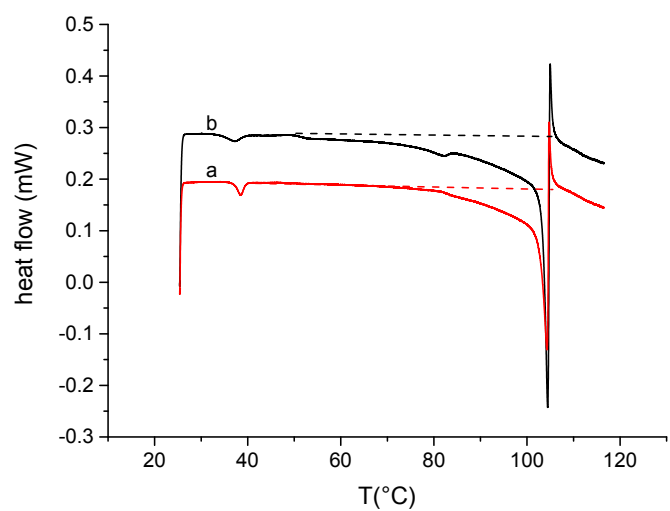

**Figure 11** Heat flow traces of *O*-palmitoyl-dextran/α-CD platelets in water (a) and in emulsion between water and silicone oil (b).

It is shown in Figure 11 that the temperature of the melting peak is unchanged with silicone oil, while there is a slight tendency to dissolve the platelets at lower temperatures, starting at approximately 50°C, instead of 70°C in water. The dissolution may be enhanced by the adsorption of the platelets at the oil droplets surface. The inclusion complex platelets lower the interfacial tension between silicone oil and water and the alkyl chain inclusions can be easily directed towards the oil phase. The areas below the base lines are for *O*-palmitoyl-dextran/α-CD in water between 30°C and 104.8°C the area is 1.17 J and in emulsion with silicone oil, 1.4 J. The difference between the areas calculated in the two cases is mainly related to this phenomenon of dissolution. However, the heating ramps end at 120°C, and all inclusion complexes are not dissolved, in both cases, as the heat flow does not return to base line. We have no interpretation for the exothermal peak (at 105°C) following the endothermal melting peak. There seems to be an additional contribution to the endothermal signal between 110 and 120°C. We may assume that the total heat of dissolution of the platelets should be unchanged in emulsion.

### 4.3.3. Discussion: Pickering emulsions with micro-platelets

This preliminary investigation shows that the platelets, which are water soluble, tend to aggregate spontaneously when they are kept at room temperature in the preparation solution. A network of small platelets (1μm) of colloidal size forms in water. When diluted 10 times with a mild agitation the "strings" of particles do not completely disperse and form small patches which provide an additional stability to the Pickering emulsion with silicone oil. Emulsions were not prepared in standardized conditions to allow a precise investigation of their time stability against droplet coalescence and sedimentation of the aggregates. The droplet size distribution is not well defined. The droplets observed were generally small (20μm diameter). It is shown that the small platelets have interfacial properties and are preferentially adsorbed at the droplet surface. The melting transitions temperature of the platelets in emulsions or dispersed in water are similar, showing a cooperative transition. However the experiments suggest that the contact between the platelets and the oil droplets enhances the dissociation or de-threading of the inclusions from the platelets. This interaction is different from the case when the oil can be threaded inside the CD molecules and destabilizes the platelets by exchanging the guest molecules (Figures 7 and 8). The particle size matters in the stability of Pickering emulsions. Mathapa and Paunov [21] studied cyclodextrin stabilized emulsions, where inclusions grow at interfaces between oil (*n*-tetradecane) and α-CD in solution in water. They kept the CD concentration constant (10 mM or 9.7g/L, which is 10 times more diluted than $\alpha$-CD aqueous solutions in the present work). The oil-water mixture was homogenized using an UltraTurrax operating at 11 000 rpm for a period of 20 seconds. At very small oil volume fractions, they obtained only precipitates at the bottom of the sample tubes. They obtained Pickering emulsions with rod-like particles which are more stable when the microrods have lengths <10µm, while the longer ones >20µm form a bottom layer. Particles with high aspect ratios are particularly efficient in stabilization of emulsions, more efficient than spherical ones [32].

### 5.4    Conclusion

Pickering emulsions based on inclusion complexes with $\alpha$-cyclodextrin were investigated in this paper with the aim to analyze their thermal stability in water and in emulsions at the interfaces with different oils. This work evidences for the first time the interplay between substitution of guest molecules at room temperature and dissolution/melting of inclusion complexes.

The enthalpy of dissolution or melting and de-threading of the crystalline inclusion complexes was determined for the first time. Contrary to previous investigations limited to dilute cyclodextrin inclusions [29–31], the enthalpy contains the large contribution arising from the crystal stability. Measured for the first time, very high meting temperatures were found, near to or higher than 100 °C.

When inclusion complexes were obtained from grafted polysaccharides and $\alpha$-CD, according to the method reported by Bouchemal et al[23], they exhibit platelet shapes with various sizes. It is shown in this paper that emulsions can be easily prepared with dispersions of such platelets, with sizes <1µm in water. Pickering emulsions with platelet like shapes were less reported in the literature compared to micro rod shapes in particular for $\alpha$-CD inclusions[21]. Based on micro DSC experiments, two different cases were clearly identified according to oil molecules: either i) the oil (such as silicone oil) is unable to form inclusion complexes with $\alpha$-CD, then the platelets keep their structure and (almost) their initial thermal stability, emulsions are stable in time and temperature; or ii) if the oil molecules could be guest molecules of the complexes, they readily substitute to the existing ones by simple agitation at room temperature and the platelets lose their initial composition and their thermal stability. The substitution possibly proceeds by a diffusion mechanism at room temperature without melting of crystals. When inclusion complexes are obtained with grafted polysaccharides it is explicitly shown that substitution of the grafted alkyl chains with fatty acids takes place rapidly. These interesting properties may be important in the pharmaceutical applications designed for platelets and may be at the origin of specific effects modifying their therapeutic benefits. The fact that substitution of inclusions is likely to take place spontaneously is an important observation. Other guest molecules belonging to the biological environment may exchange with those present in the platelets. Exchanges could be also influenced by other molecules present in formulations.

### Acknowledgements:

We are grateful to Dr. Suzanne Bolte from Institut de Biologie Paris-Seine (IBPS) for performing CM observations and for very helpful discussions.

### References

[1]    S.U. Pickering, Cxcvi.—emulsions, J. Chem. Soc. Trans. 91 (1907) 2001–2021.

[2]    E. Dickinson, Food Colloids  -  An  Overview, Colloids Surf. 42 (1989) 191–204.

[3]    E. Dickinson, Hydrocolloids as emulsifiers and emulsion stabilizers, Food Hydrocoll. 23 (2009) 1473–1482. doi:10.1016/j.foodhyd.2008.08.005.

[4]    E. Dickinson, Use of nanoparticles and microparticles in the formation and stabilization of food emulsions, Trends Food Sci. Technol. 24 (2012) 4–12. doi:10.1016/j.tifs.2011.09.006.

[5]   M.V. Tzoumaki, T. Moschakis, V. Kiosseoglou, C.G. Biliaderis, Oil-in-water emulsions stabilized by chitin nanocrystal particles, Food Hydrocoll. 25 (2011) 1521–1529. doi:10.1016/j.foodhyd.2011.02.008.

[6]   B.P. Binks, S.O. Lumsdon, Catastrophic Phase Inversion of Water-in-Oil Emulsions Stabilized by Hydrophobic Silica, Langmuir. 16 (2000) 2539–2547. doi:10.1021/la991081j.

[7]   B.P. Binks, J.H. Clint, Solid Wettability from Surface Energy Components: Relevance to Pickering Emulsions, Langmuir. 18 (2002) 1270–1273. doi:10.1021/la011420k.

[8]   B.P. Binks, J. Philip, J.A. Rodrigues, Inversion of Silica-Stabilized Emulsions Induced by Particle Concentration, Langmuir. 21 (2005) 3296–3302. doi:10.1021/la046915z.

[9]   B.P. Binks, C.P. Whitby, Silica Particle-Stabilized Emulsions of Silicone Oil and Water: Aspects of Emulsification, Langmuir. 20 (2004) 1130–1137. doi:10.1021/la0303557.

[10]  B.R. Midmore, Preparation of a novel  silica-stabilized oil/water emulsion, Colloids Surf. Physicochem. Eng. Asp. 132 (1998) 257–265.

[11]  C.P. Whitby, D. Fornasiero, J. Ralston, Effect of adding anionic surfactant on the stability of Pickering emulsions, J. Colloid Interface Sci. 329 (2009) 173–181. doi:10.1016/j.jcis.2008.09.056.

[12]  D.E. Tambe, M.M. Sharma, Factors controlling the stability of colloid stabilized emulsions, J Colloid Interface Sci. 162 (1994) 1–10.

[13]  A. Harada, M. Kamachi, Complex Formation between Poly(ethy1ene glycol)  and a-Cyclodextrin, Macromolecules. 23 (1990) 2823–2824.

[14]  A. Harada, J. Li, M. Kamachi, Preparation and properties of inclusion complexes of polyethylene glycol with. alpha.-cyclodextrin, Macromolecules. 26 (1993) 5698–5703.

[15]  A. Harada, J. Li, M. Kamachi, Formation of Inclusion Complexes of Monodisperse Oligo (ethylene glycol) s with. alpha.-Cyclodextrin, Macromolecules. 27 (1994) 4538–4543.

[16]  J. Li, A. Harada, M. Kamachi, Bull Chem Soc Jpn. 67 (1994) 2808–2818.

[17]  K. Hashizaki, T. Kageyama, M. Inoue, H. Taguchi, H. Ueda, Y. Saito, Study on Preparation and Formation Mechanism of n-Alkanol/Water Emulsion Using. ALPHA.-Cyclodextrin, Chem. Pharm. Bull. (Tokyo). 55 (2007) 1620–1625.

[18]  M. Inoue, K. Hashizaki, H. Tagushi, Saito, Preparation and characterization of n-alkane water emulsion stabilized by cyclodextrin, J Oleo Sci. 58 (2009) 85–90.

[19]  I. Sanemasa, T. Osajima, T. Deguchi, Association C5-C9 normal alkanes with cyclodextrins in aqueous medium, Bull Chem Soc Jpn. 63 (1990) 2814–2819.

[20]  A.K.F. Dyab, V.N. Paunov, Particle stabilised emulsions studied by WETSEM technique, Soft Matter. 6 (2010) 2613. doi:10.1039/c0sm00032a.

[21]  B.G. Mathapa, V.N. Paunov, Cyclodextrin stabilised emulsions and cyclodextrinosomes, Phys. Chem. Chem. Phys. 15 (2013) 17903. doi:10.1039/c3cp52116h.

[22]  B.G. Mathapa, V.N. Paunov, Self-assembly of cyclodextrin–oil inclusion complexes at the oil–water interface: a route to surfactant-free emulsions, J. Mater. Chem. A. 1 (2013) 10836. doi:10.1039/c3ta12108a.

[23]  K. Bouchemal, Microparticles and nanoparticles made of hydrophobized polysaccharide and alpha-cyclodextrin., WO/2013/150193, 2013.

[24]  A. Galus, J.-M. Mallet, D. Lembo, V. Cagno, M. Djabourov, H. Lortat-Jacob, K. Bouchemal, Hexagonal-shaped chondroitin sulfate self-assemblies have exalted anti-HSV-2 activity, Carbohydr. Polym. 136 (2016) 113–120. doi:10.1016/j.carbpol.2015.08.054.

[25]  H. Okumura, Y. Kawaguchi, A. Harada, Preparation and Characterization of Inclusion Complexes of Poly(dimethylsiloxane)s with Cyclodextrins, Macromolecules. 34 (2001) 6338–6343. doi:10.1021/ma010516i.

[26]  E. Specogna, K.W. Li, M. Djabourov, F. Carn, K. Bouchemal, Dehydration, Dissolution, and Melting of Cyclodextrin Crystals, J. Phys. Chem. B. 119 (2015) 1433–1442. doi:10.1021/jp511631e.

[27]  D. Duchêne, A. Bochot, S.-C. Yu, C. Pépin, M. Seiller, Cyclodextrins and emulsions, Int. J. Pharm. 266 (2003) 85–90. doi:10.1016/S0378-5173(03)00384-3.

[28]  A. Bochot, L. Trichard, G. Le Bas, H. Alphandary, J.L. Grossiord, D. Duchêne, E. Fattal, α-Cyclodextrin/oil beads: An innovative self-assembling system, Int. J. Pharm. 339 (2007) 121–129. doi:10.1016/j.ijpharm.2007.02.034.

[29]  G. Castronuovo, V. Elia, D. Fessas, F. Velleca, G. Viscardi, Thermodynamics of the interaction of a-cyclodextrin with monocarboxylic acids in aqueous solutions: a calorimetric study at 25 °C, Carbohydr. Res. 287 (1996) 127–138.

[30]  G. Castronuovo, V. Elia, F. Valleca, Hydrophilic groups determine preferential configurations in aqueous solutions. A calorimetric study of monocarboxylic acids and monoalkylamines at 298.15 K, Thermochim. Acta. 291 (1997) 21–26.

[31]  G. Castronuovo, V. Elia, M. Niccoli, F. Velleca, G. Viscardi, Role of the functional group in the formation of the complexes between α-cyclodextrin and alkanols or monocarboxylic acids in aqueous solutions. A calorimetric study at 25° C, Carbohydr. Res. 306 (1998) 147–155.

[32]  E. Tervoort, A.R. Studart, C. Denier, L.J. Gauckler, Pickering emulsions stabilized by in situ grown biologically active alkyl gallate microneedles, RSC Adv. 2 (2012) 8614. doi:10.1039/c2ra21253f.

## Captions

**Figure 1** (a) Emulsion between hexanoic acid and water solution containing 10wt.% $\alpha$-CD after 30 min agitation at room temperature. The emulsion is a white phase and the oil is a pink fluid layer on the top. (b) The vial was put in the oven at different temperatures, up to 75°C.

**Figure 2** Thermal behavior of inclusion complexes formed in emulsions of hexanoic acid with aqueous solutions: a) no $\alpha$-CD in the water; b) The aqueous solution contains $\alpha$-CD, (10 wt.%) and was mixed with hexanoic acid 30 min before measuring; c) After 3 days mixing, more stable inclusions are formed and a distinct crystalline phase appears which melts at the temperature $T_m$=72.7°C; d) After 5 days emulsification, the melting peak position is unchanged. The total area under the base line increases. The dotted lines indicate the base lines of the heat flow.

**Figure 3** Octanoic acid inclusions formed at the interface between water and octanoic acid without agitation (a); after shaking rapidly by hand at 100°C (b). The vial was put in the oven at different temperatures, up to 100°C. The oil phase is the pink color; the inclusion complex form large, macroscopic crystals which sediment at the bottom.

**Figure 4** Thermal behavior of inclusion complexes formed in emulsions of octanoic acid with $\alpha$-CD in aqueous solutions (10wt%) (3 days under agitation before measuring). The heat flow trace shows a clear melting peak of the crystalline inclusion complexes between $\alpha$-CD and octanoic acid at $T_m$=92°C.

**Figure 5** Heat flow traces of emulsions formed with olive oil and water containing 10wt% $\alpha$-CD. The O/W fraction was varied by weight from top to bottom: 10/90; 25/75; 40/60wt%. Emulsions were magnetically agitated during 24h prior to µDSC experiments.

**Figure 6** X ray diffraction patterns of inclusion complexes collected after centrifugation of the emulsions with different oils. From top to bottom: octanoic acid, hexanoic acid and olive oil inclusion complexes with $\alpha$-CD.

**Figure 7** Heat flows: a) (red line) Platelets composed of *O*-palmitoyl-dextran/$\alpha$-CD inclusion complexes (dexPA/$\alpha$-CD) in water; b) (green line) platelets in water and emulsion with hexanoic acid

(HA); c) (blue line) emulsion between hexanoic acid and α-CD in water. The platelets in emulsion with hexanoic acid lost their thermal stability.

**Figure 8** Heat flow trace for *O*-palmitoyl-dextran/α-CD micro-platelets (dexPA/α-CD) in water, emulsified with octanoic acid (OA). The heat flow is totally changed in comparison with Figure 7.

**Figure 9** Platelets of *O*-palmitoyl-dextran/α-CD in undiluted suspension, in emulsion 25/75wt% with silicone oil. Left: transmission mode; right reflection mode (bar=10μm).

**Figure 10** Platelets of *O*-palmitoyl-amylopectin/α-CD diluted 10 times in emulsions with 10/90wt.% silicone oil in transmission (left) and reflection (right) modes (bar = 10μm).

**Figure 11** Heat flow traces of *O*-palmitoyl-dextran/α-CD platelets in water (a) and in emulsion between water and silicone oil (b).

## Table

| Hexanoic acid emulsion | | Octanoic acid emulsion | Olive oil emulsion | |
|---|---|---|---|---|
| 30 min | 1.55 ± 0.10 J | - | 10/90 | 0.31 ± 0.10 J |
| 3 days | 2.38 ± 0.10 J | 2.86 ± 0.10 J | 25/75 | 1.38 ± 0.10 J |
| 5 days | 2.63 ± 0.10 J | - | 40/60 | 1.57 ±0.10 J |
| **|ΔH| /mole α-CD** | **39 ± 1 kJ/mole** | **>36 ± 1 kJ/mole** | **Max ~32 ± 1 kJ/mole** | |

**Table 1.** Enthalpies of dissolution or melting and de-threading of inclusions self-assembled in crystalline forms with different oils.

# Supporting data

This section describes the preparation and the physico-chemical characterization of *O*-palmitoyl-dextran or *O*- *O*-palmitoyl-amylopectin and their corresponding platelets obtained from the inclusion complexes between hydrophobically-modified polysaccharides and α-cyclodextrin.

# 1. Preparation and the physico-chemical characterization of platelets

## 1.1. Preparation of O-palmitoyl-dextran or O-palmitoyl-amylopectin

Dextran, amylopectin, palmitoyl chloride, pyridine and dry dimethylformamide (DMF, 99.8%) were from Sigma (Saint-Quentin Fallavier, France). Pullulan was from Fluka. Absolute ethanol and diethylether (Carlo Erba) were from Carlo Erba, France. α-cyclodextrin was from Cyclolab (Budapest, Hungary).

*O*-palmitoyl-dextran, or *O*-palmitoyl-amylopectin were prepared by dissolving dextran, pullulan or amylopectin (5 g) in 55 mL of dry DMF. The mixture was heated to 60°C under continuous magnetic stirring. Then, 5 mL of dry pyridine and 1.2 mL of dry DMF containing 7 g of palmitoyl chloride were introduced to the resulting solution. After 2 hours at 60°C, and 1 hour at room temperature, the mixture was poured onto 350 mL ethanol. Precipitates were collected and washed with 400 mL of ethanol and, then, with 300 mL of diethylether using a Buchner filter. The solid materials were dried under vacuum at 50°C for 2 h.

## 1.2. Chemical characterization of O-palmitoyl-dextran or O-palmitoyl-amylopectin

The obtained *O*-palmitoyl-dextran or *O*-palmitoyl-amylopectin were then characterized by using Attenuated total reflectance-Fourier transform infrared (ATR-FTIR) spectroscopy by a spectrometer (FT/IR-4100, JASCO) operating at 4 cm$^{-1}$ resolution. Fifty scans were accumulated in each run and referred to air. The ATR sampling device utilized a diamond internal reflection element embedded into a ZnSe support/focusing element in a single reflection configuration. The resultant spectrum over the range of 4000–400 cm$^{-1}$ was analyzed using the IR Protein Secondary Structure Analysis program (JASCO Co).

## 1.3. Platelet physico-chemical characterization

The hydrodynamic diameter of the platelets was determined at 25°C by quasi-elastic light scattering using a Zetasizer Nanoseries Nano-ZS (Malvern Instruments, France). The scattered angle was fixed at 173° and 30 μL of each sample was diluted in 1 mL of MilliQ$^{®}$ water. Transmission Electron Microscopy (TEM) images were obtained using the transmission electron microscope of 60 kV Jeol 1400 (Imagif, Gif sur Yvette, France). For this, 1 μL of the platelet suspension was diluted in 29 μL of MilliQ$^{®}$ water. Then, 3 μL of this dilution are placed on a grid. After 5 minutes of drying, the grid is inserted into the microscope to view the sample.

## 1.4. Hydrophobically-modified polysaccharide characterization

Infrared spectroscopy of esterified dextran or amylopectin showed stretching vibration bands at 1666 to 1796 cm$^{-1}$ corresponding to C=O ester bonds (Figures 1-2 and Table 1). The vibrations of C-H stretching of CH$_2$ and CH$_3$ groups of palmitic acid are around 2848-2954 cm$^{-1}$. –CH$_2$ deformation bands of palmitic acid were reported at 1464-1771 cm$^{-1}$.

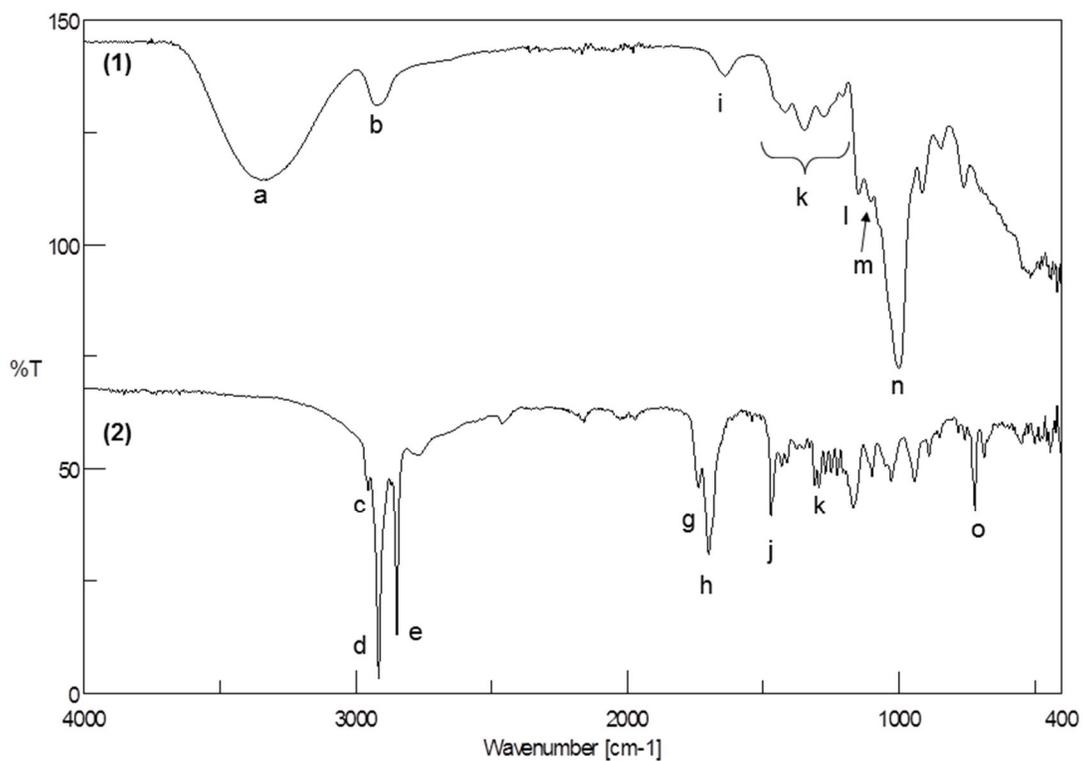

**Figure 1** ATR-FTIR spectra of dextran (**1**) and *O*-palmitoyl-dextran (**2**).

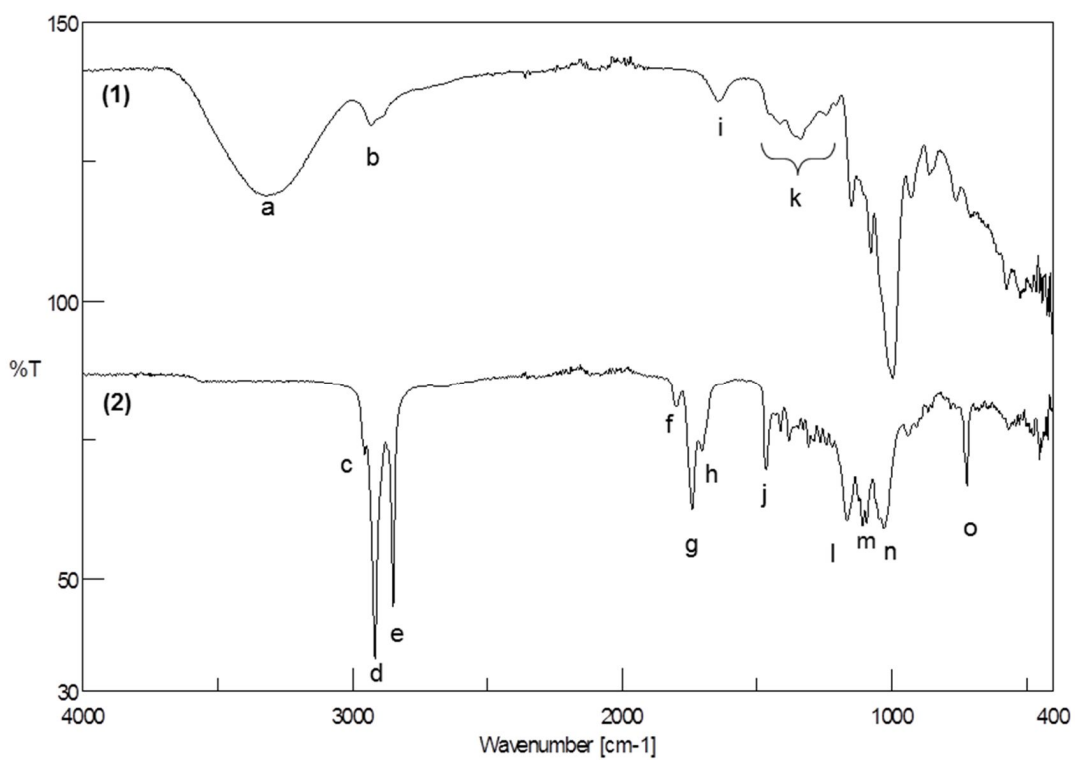

**Figure 2** ATR-FTIR spectra of amylopectin (**1**) and *O*-palmitoyl-amylopectin (**2**).

**Table 1:** ATR-FTIR band assignments of native and esterified dextran and amylopectin.

| Code | Dextran | *O*-palmitoyl-dextran | Amylopectin | *O*-palmitoyl-amylopectin | Assignments |
|------|---------|-----------------------|-------------|---------------------------|-------------|
| a | 3447 | | 3316 | | Stretching vibrations of O-H bounds of dextran, or amylopectin |
| b | 2926 | | 2929 | | Stretching vibrations of O-H bounds of -CH$_2$ groups of dextran, or amylopectin |
| c | | 2953 | | 2954 | Stretching vibrations of C-H bounds of -CH$_2$ and -CH$_3$ groups of palmitic acid |
| d | | 2915 | | 2917 | |
| e | | 2848 | | 2849 | |
| f | | | | 1796 | -C=O stretching carbonyl of ester groups |
| g | | 1738 | | 1739 | |
| h | | 1700 | | 1702 | |
| i | 1638 | | 1644 | | δ(O-H) |
| j | | 1471 | | 1464 | -CH$_2$-, deformation of palmitic acid |
| k | 1416-1207 | | 1416-1243 | | C-OH in plane bend |
| l | 1149 | 1168 | 1149 | 1165 | Stretching of C-O and CH$_2$-OH. δ(CH$_2$) |
| m | 1103 | 1099 | 1077 | 1107 | |
| n | 999 | 1027 | 997 | 1027 | |
| o | | 721 | | 720 | C-C=O bend |

## 1.5. Platelet size characterization

Without α-CD, hydrophobically-modified polysaccharides are insoluble in water and aggregates were formed at a concentration of 1wt%. Platelets were formed after 72h mixing with α-CD aqueous solution. Dynamic light scattering measurements (Table 2) showed that platelet mean hydrodynamic diameters were in the micrometer range whatever the polysaccharide used. Noteworthy, with lower amounts of α-CD (5 and 7.5wt%), aggregates were formed. Observation of platelet suspensions using TEM revealed well-structured, hexagonal-shaped particles (Figure 3).

**Table 2:** Mean hydrodynamic diameter (D$_h$) determinations of platelets composed of *O*-palmitoyl-dextran/α-CD, and *O*-palmitoyl-amylopectin/α-CD inclusion complexes. The weight ratio between hydrophobically-modified polysaccharides and α-CD was 1/10 wt%. n=3.

| Platelet composition | D$_h$ (nm) |
|----------------------|-----------|
| *O*-palmitoyl-dextran/α-CD | 2674 ± 805 |
| *O*-palmitoyl-amylopectin/α-CD | 2087 ± 408 |

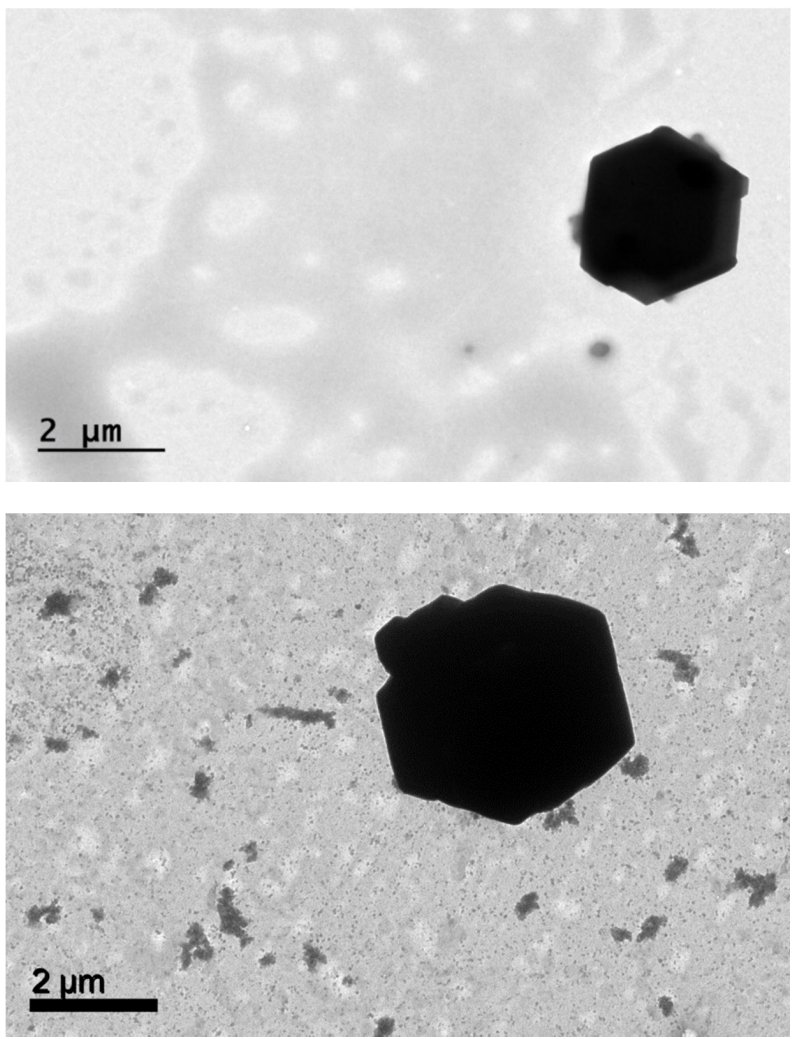

**Figure 3** TEM image of platelets composed of *O*-palmitoyl-dextran/α-CD (top), and *O*-palmitoyl-amylopectin/α-CD inclusion complexes (bottom).

## 2. Emulsions between octanoic acid and α-CD in solution

Emulsion was first agitated during 30min with a magnetic stirrer at room temperature. In Figure 5 at 40°C the emulsion is at the bottom (white) and the oil forms a thin layer at the top (pink color). At 60°C three layers are seem: at the bottom, the denser phase is the "emulsion" (white) containing the inclusion complex crystals and water, the intermediate phase is the aqueous phase with the small inclusion complexes in suspension (turbid phase), the top phase is the oil. At 75°C, the small inclusion complex crystals dissolve progressively and the intermediate water phase becomes transparent; the oil forms a film at the surface, but with some inclusion complexes which persist at the interface with the water, forming a white layer. It is seen that water degases above 80°C and small bubbles together with oil droplets rise at the surface and accumulate as a separate layer, forming a kind of light "foam" maybe due to the interfacial properties of the inclusions. This top layer is persistent at 100°C, while the crystals at the bottom have melted. The vial was not agitated and it is possible that there is a kinetic effect due to the concentration gradient of α-CD inside the aqueous phase or a thermal gradient which retards the dissolution of the top layer at 100°C.

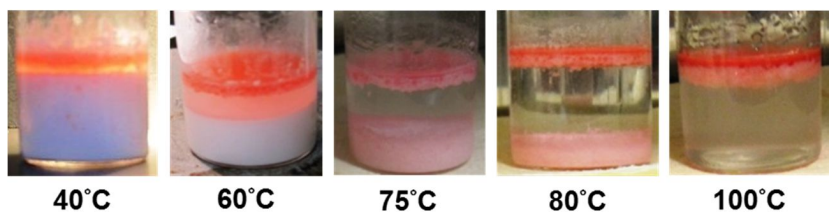

**Figure 5:** Emulsions between octanoic acid and water solution with 10% α-CD after 30 min agitation. The emulsion is a white phase and the oil is a pink fluid layer on the top. The vial was put in the oven at different temperatures, up to 100°C.

## 3. Confocal microscopy images of micro platelets and silicone oil droplets

Figure 6 left hand, is the image in transmission mode of the suspension of *O*-palmitoyl-amylopectin/α-CD platelets as prepared in water with a droplet of silicone oil in the middle. The bar is 10μM. One can see the dense suspension of platelets with a large size distribution between 1-5μm and a few large hexagonal particles. The droplet is covered with the smallest platelets, well identified on the periphery of the drop. The largest platelets are excluded from the drop surface.

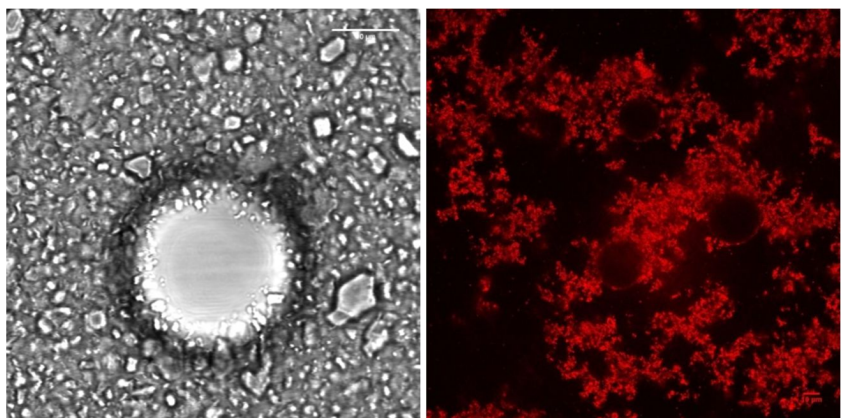

**Figure 6** Pickering emulsion between *O*-palmitoyl-amylopectin/α-CD platelets in their initial suspension and silicone oil droplets. Left: transmission mode; right; reflection mode (bar=10μm).

Figure 7 right hand, shows in reflection mode a collection of small droplets of silicone oil with the suspension of platelets of *O*-palmitoyl-amylopectin/α-CD aggregated as a continuous network.

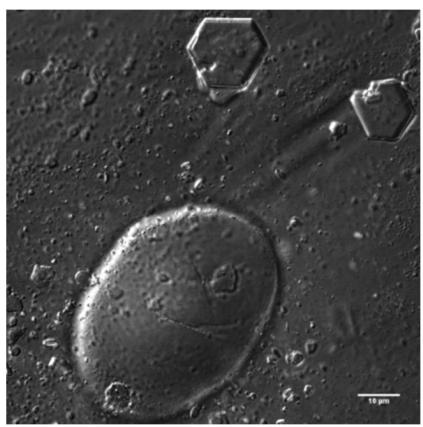

**Figure 7** *O*-palmitoyl-amylopectin/α-CD diluted suspension in emulsion 50/50wt% with silicone oil (bar 10μm).

Figure 7 shows another image of *O*-palmitoyl-amylopectin/$\alpha$-CD diluted suspension of platelets in emulsion 50/50wt% with silicone oil where an isolated large droplet (50µm) is covered with small platelets (1µm), while large hexagons (10µm) and of intermediate size lie flat, away from the droplet, in the water phase. The method of synthesis of the platelets creates polydisperse platelets. This image shows that large platelets are seen by confocal microscopy without any specific treatment before observation, contrary to TEM or SEM techniques.